**Comment on "Large enhancement in high-energy photoionization of Fe XVII and missing continuum plasma opacity"**

Recent R-matrix calculations claim to produce a significant enhancement in the opacity of Fe XVII due to atomic core excitations [1] and assert that this enhancement is consistent with recent measurements of higher-than-predicted iron opacities [2]. This comment shows that the standard opacity models [3-7] which have already been directly compared with experimental data [2,7] produce photon absorption cross-sections for Fe XVII that are effectively equivalent to the R-matrix opacities reported in [1]. Thus, the new R-matrix results cannot be expected to significantly impact the existing discrepancies between theory and experiment because they produce neither a "large enhancement" nor account for "missing continuum plasma opacity" relative to standard models.

All models that satisfy the f-sum rule [7] and include the same initial and final electronic configurations can be expected to produce similar opacities (*e.g.* [8]). This is demonstrated in Fig.1, which compares calculated opacities for Fe XVII from five standard models to the R-matrix and OP results from [1]. The models have been restricted to the Fe XVII ion and normalized to a 0.195 abundance but are otherwise the same as those previously published [2,7]. Both R-matrix and standard models include spectral features associated with autoionizing states that are evident in measured data but neglected in OP [1,9]. Thus the opacity enhancements of R-matrix over OP reported in [1] illustrate the deficiencies of OP rather than the merits of R-matrix.

Table 1 gives relative opacities for Fe XVII to help quantify the similarities between R-matrix and standard models and their mutual differences with measurements. Both R-matrix and standard models yield larger total Rosseland mean opacities than OP, confirming the importance of transitions missing in OP. However, the Rosseland weighting function peaks near 17 Å while the most profound discrepancies between theory and experiment are in the 7 - 9 Å monochromatic continuum range. Here, the average opacities from all models (as well as cold reference opacities [10]) are significantly smaller than the experimental data. In this critical range, R-matrix is smaller even than OP. Thus the results reported in [1] appear unlikely to resolve this discrepancy between theory and experiments.

C. Blancard[1], J. Colgan[2], Ph. Cossé [1], G. Faussurier[1], C. J. Fontes[2], F. Gilleron[1], I. Golovkin[3], S. B. Hansen[4*], C. A. Iglesias[5], D. P. Kilcrease[2], J. J. MacFarlane[3], R.M. More[6], J.-C. Pain[1], M. Sherrill[2], and B. G. Wilson[5]

[1] *Commissariat a l'Energie Atomique et aux Energies Alternatives, F-91297 Arpajon, France*
[2] *Los Alamos National Laboratories, Bikini Atoll Road, Los Alamos, NM 87545, USA*
[3] *Prism Computational Sciences, 455 Science Drive, Suite 140, Madison, WI 53711, USA*
[4] *Sandia National Laboratories, 1515 Eubank SE, Albuquerque, NM 87185, USA*
[5] *Lawrence Livermore National Laboratories, P.O. Box 808, Livermore, CA 94550, USA*
[6] *National Institute for Fusion Science, Toki, Gifu, Japan (ret.)*

[*]Corresponding author: S. B. Hansen, e-mail sbhansen@sandia.gov

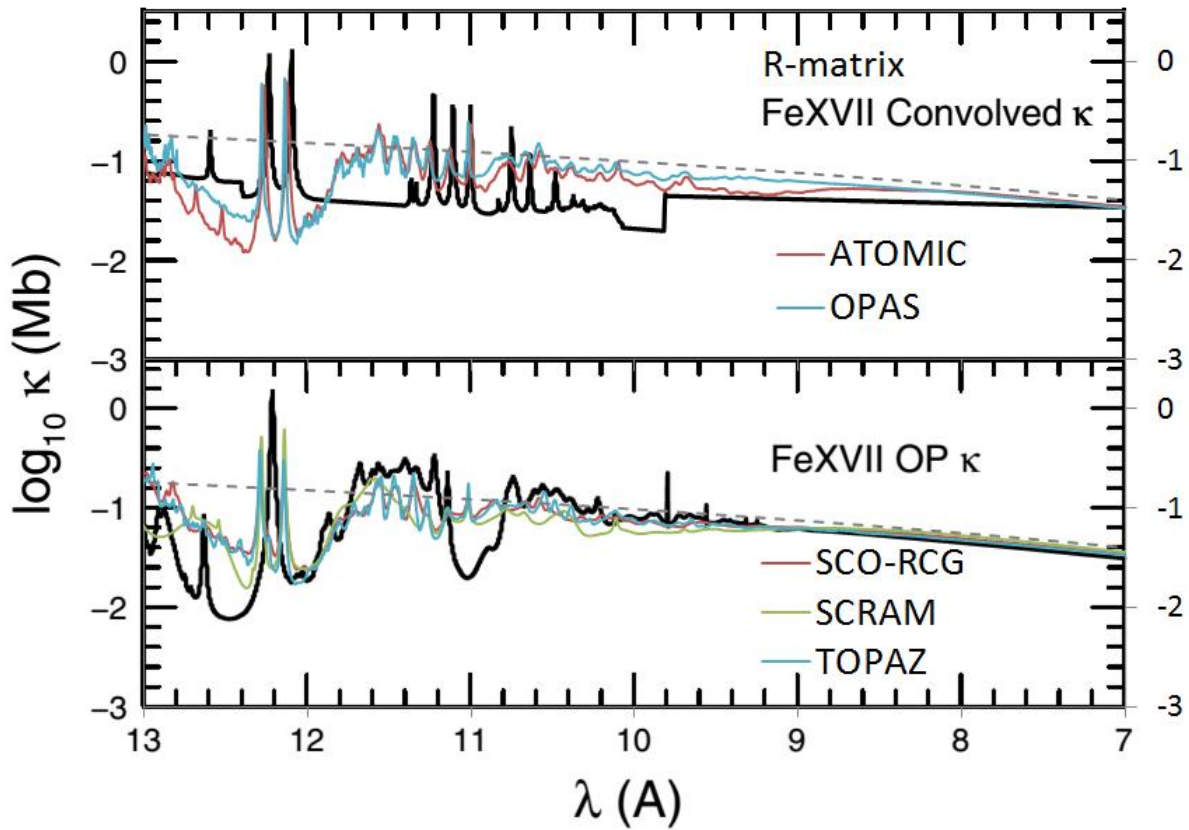

**Fig.1** (Color online; adapted from Fig. 5 of Ref. [1]) Opacities of FeXVII at a temperature of $2.1 \times 10^6$ K, free electron number density of $3.1 \times 10^{22}$ cm$^{-3}$, and abundance of 0.195. Dashed lines are 0.195 x the cold reference opacity [10], representing a fully occupied L-shell.

**Table 1** Rosseland mean opacities $\kappa_R$ of Fe XVII normalized to the OP value demonstrate that both R-matrix and standard models are significantly larger than OP. Average Fe XVII opacities $<\kappa>$ in the 7- 9 Å continuum region normalized to experimental data [2] show deficits in all models.

| Source | $\kappa_R$ (total) relative to OP [1] | $<\kappa>$ (7 − 9 Å) relative to experiment [2] |
|---|---|---|
| OP [1] | 1.00 | 0.59* |
| R-matrix [1] | 1.35 | 0.52* |
| ATOMIC [3] | 1.32 | 0.60 |
| OPAS [4] | 1.55 | 0.62 |
| SCO-RCG [5] | 1.37 | 0.65 |
| SCRAM [6] | 1.27 | 0.68 |
| TOPAZ [7] | 1.21 | 0.62 |
| Cold [10] | | 0.74 |
| Experiment [2] | | 1.00 |

* Estimated from Fig. 5 of Ref. [1].

**Acknowledgements:** The work of CAI and BGW was performed under the auspices of the U.S. Department of Energy by Lawrence Livermore National Laboratory under Contract DE-AC52-07NA27344. The work of JC, CJF, DPK, and MS was performed at The Los Alamos National Laboratory, operated by Los Alamos National Security, LLC, for the NNSA of the US DOE under contract number DE-AC5206NA25396. SBH was supported by the U.S. Department of Energy, Office of Science Early Career Research Program, Office of Fusion Energy Sciences. Sandia National Laboratories is a multiprogram laboratory managed and operated by Sandia Corporation, a wholly owned subsidiary of Lockheed Martin Corporation, for the U.S. Department of Energy's National Nuclear Security Administration under Contract No. DE-AC04-94AL85000.